\documentclass[10pt,aps,twocolumn,prd,superscriptaddress,showpacs,nofootinbib,noshowkeys,floatfix,preprintnumbers]{revtex4}
\usepackage{times}
\usepackage{graphics,graphicx}
\usepackage{amsmath, amssymb}
\usepackage{multirow}
\usepackage{longtable}
\usepackage{color}

\usepackage[normalem]{ulem}

\begin{document}
\title{Polyakov loop fluctuations and  deconfinement  in the limit of heavy quarks
\\
}
\date{\today}
\date{\today}
\author{Pok Man Lo}
\affiliation{GSI, Helmholzzentrum f\"{u}r Schwerionenforschung,
Planckstr. 1, D-64291 Darmstadt, Germany}
\author{Bengt Friman}
\affiliation{GSI, Helmholzzentrum f\"{u}r Schwerionenforschung,
Planckstr. 1, D-64291 Darmstadt, Germany}
\author{Krzysztof Redlich}
\affiliation{Institute of Theoretical Physics, University of Wroclaw,
PL-50204 Wroc\l aw, Poland}
\affiliation{Extreme Matter Institute EMMI, GSI,
Planckstr. 1, D-64291 Darmstadt, Germany}
\begin{abstract}
 {
We explore the influence of heavy quarks on the deconfinement phase transition in an effective model for gluons interacting with dynamical quarks in color SU(3). With decreasing quark mass, the strength of the explicit breaking of the Z(3) symmetry grows and the first-order transition ends in a critical endpoint. The nature of the critical endpoint is examined by studying the longitudinal and transverse fluctuations of the Polyakov loop, quantified by the corresponding susceptibilities. The longitudinal susceptibility is enhanced in the critical region, while the transverse susceptibility shows a monotonic behavior across the transition point. We investigate the dependence of the critical endpoint on the number of quark flavors at vanishing and finite quark density. Finally we confront the model results with lattice calculations and discuss a possible link between the hopping parameter and the quark mass.
   }
\end{abstract}


\maketitle

\section{Introduction}

In the limit of infinitely heavy quarks, deconfinement in the SU(3) gauge theory is associated with the spontaneous breaking of the Z(3) center symmetry \cite{Greensite, tHooft, Yaffe, McLerran, Polyakov}. The phase transition is of first-order and is therefore stable against small explicit symmetry breaking \cite{Binder}. The presence of dynamical quarks breaks the center symmetry explicitly, with a strength that increases as the quark mass decreases. It is thus expected that the transition remains discrete in the heavy quark region, and becomes continuous at some critical value of quark mass \cite{GK}. This defines the critical endpoint (CEP) for the deconfinement phase transition.

The relevant observables to study deconfinement are the Polyakov loop and its susceptibilities. Recently these quantities were computed on the lattice within SU(3) pure gauge theory \cite{pmlo1}.
In particular, the ratio of the transverse and longitudinal Polyakov loop   susceptibility was shown to  exhibit a $\theta$-like behavior  at the critical temperature $T_d$, with almost no dependence on     temperature on either side of the transition. This feature makes  such   ratios   ideal  probes  of  deconfinement.

The influence of the dynamical light quarks on these susceptibilities has been studied in 2- and (2+1)-flavor QCD  \cite{2+1QCD, pmlo2}. 
The resulting susceptibility ratios are considerably smoothened, reflecting the crossover nature of the transition. However, the theoretical understanding of these observables is still incomplete. It is therefore useful to explore the properties of the Polyakov loop susceptibilities for different number of flavors, as functions of the quark mass in the heavy quark region, thus bridging the gap between pure gauge theory and QCD.

The thermodynamics of SU(3) gauge theory with heavy quarks have long been studied in an effective theory \cite{GK} and on the lattice \cite{Karsch2001579, DeGrand, attig, Langelage:2009jb, Ejiri:2012wp, PhysRevD.60.034504, Aarts:2001dz, fomm, whot}. The nature of the deconfinement CEP was recently examined within the effective matrix model \cite{PhysRevD.85.114029}. One key finding of this study is that the model results for the critical point depend strongly on the structure of the phenomenological gluon potential.

In this work we reexamine the phase structure of the deconfinement transition for heavy quarks. We formulate an  effective theory where the gluon potential is constructed using the pure lattice gauge theory results on the equation of state and on fluctuations of the Polyakov loop \cite{pmlo2}. The contribution from heavy quarks is described by the one loop thermodynamic potential of fermions coupled to a background gluon field \cite{meisinger}.

To study the deconfinement CEP, it is essential that the phenomenological gluon potential reproduces not only the equation of state and the Polyakov loop, but also the pure gauge theory results for fluctuations of the Polyakov loop. Thus, we employ the effective Polyakov loop potential of Ref.\ \cite{pmlo2}, where both the location of minimum and the curvature are constrained by lattice results. This distinguishes this work from previous studies \cite{PhysRevD.85.114029}.

We focus on the behavior of the Polyakov loop susceptibilities near the CEP at vanishing and finite quark chemical potential. We also investigate the quark mass dependence of the critical temperature of the CEP for different number of flavors. We find that the longitudinal Polyakov loop susceptibility is strongly enhanced in the critical region, and hence, can  be used to probe the location of the CEP. On the other hand, the transverse fluctuations are insensitive to the continuous phase transition, showing monotonic behavior across the critical endpoint.

For a comparison of model and lattice results, we utilize the hopping parameter expansion of the fermionic determinant. Moreover, we propose a tentative relation between the hopping parameter and the quark mass in the continuum limit.

The paper is organized as follows. In the next section, we introduce the effective model for deconfinement in the presence of heavy quarks.  We calculate the properties of different susceptibilities and locate the deconfinement critical endpoint.  In Sec. III  we  relate the effective model result and   lattice data.   In Sec. IV we present our conclusions.

\section{Modeling deconfinement in the presence of quarks }

To explore the influence of heavy quarks on the deconfinement  phase transition
in the  SU(3) gauge theory,  we consider the following  model for   the partition function \cite{GK,gross,meisinger,kr,PhysRevD.69.097502},

\begin{align}
        \label{main_partition}
Z = \int dL d\bar{L}
 \det[\hat{Q}_F] \,  e^{-\beta V \hat U_G(L, \bar{L})},
 \end{align}
\noindent where $\beta=T^{-1}$ is the inverse temperature,  $L$ is the Polyakov loop  and $\bar{L}$ its conjugate.   Furthermore, $\hat U_G$ is the Z(3) invariant Polyakov loop potential extracted from pure gauge theory,  including the contribution of the SU(3) Haar measure. The  quark contribution is represented by the  determinant  of the fermionic  matrix $\hat{Q}_F$, which  in the uniform background gluon field $A_4$, reads

\begin{align}
        \hat{Q}_F &= (-\partial_\tau + \mu - i g  A_4)  \gamma^0 + i \vec{\gamma} \cdot \nabla - m.
\end{align}
\noindent For $N_f$ degenerate flavors of quark mass $m$, we obtain the one loop expression \cite{meisinger},

\begin{align}
        \label{quark1}
        \ln \det [ \hat{Q}_F ] &=  2V\beta N_f \int \frac{d^3 k}{(2 \pi)^3} \, ( 3 E(k) +  g^+ +  g^- ),
\end{align}

\noindent where the first term is the vacuum contribution, with $E(k) = \sqrt{k^2 + m^2}$, and

\begin{align}
        g^+ = T\ln(1 + 3L
         e^{-\beta E^+} + 3
         \bar{L}
          e^{-2 \beta E^+} + e^{-3 \beta E^+})  \label{quark}
\end{align}

\noindent is the contribution of quarks coupled to the Polyakov loop,  with $ E^+ =E(k) - \mu$. The  function  $g^-$ describes the anti-quarks,  and is obtained   from Eq.\eqref{quark} by replacing $\mu \rightarrow -\mu$  and $L \leftrightarrow \bar L$.

The thermodynamic potential density $\Omega=-(T/V)\ln Z$  is obtained from Eq.\eqref{main_partition} in the  mean field approximation,

\begin{align}
        { {T^{-4}}\Omega} = U_G(L,\bar L) +U_Q(L,\bar L)\label{MF},
\end{align}
\noindent where $U_Q$ is the fermion contribution to the effective potential and the thermal averaged Polyakov loop,  $L$ and $\bar L$, satisfy the gap equations,

\begin{align}
\partial \Omega /\partial L=0, ~~~~\partial \Omega /\partial {\bar  L}=0 \label{gap}.
 \end{align}

For the pure  gluon  part $U_G$, we  employ the following phenomenological   Polyakov loop potential \cite{pmlo2},
\begin{align}\label{glue}
        {U}_G = &-\frac{1}{2} A(T)\bar{L} L + B(T) \ln M_H \nonumber \\
        &+ \frac{1}{2} C(T) ( L^3 + \bar{L}^3 ) + D(T) (\bar{L} L)^2, 
\end{align}
where the SU(3)  Haar measure  $M_H$,    is given by
\begin{align}
 M_H &= 1-6 \bar{L} L + 4 ( L^3 + \bar{L}^3 ) -3 (\bar{L} L)^2.
 \end{align}

\noindent The potential (\ref{glue})  was  constructed so as to describe the lattice data on SU(3) thermodynamics, including  fluctuations of the Polyakov loop. The gluonic potential $U_G$  yields a first-order deconfinement phase transition at the critical temperature,  $T_d= 0.27 \, {\rm GeV}$.

\begin{figure}[t]
 \includegraphics[width=3.275in]{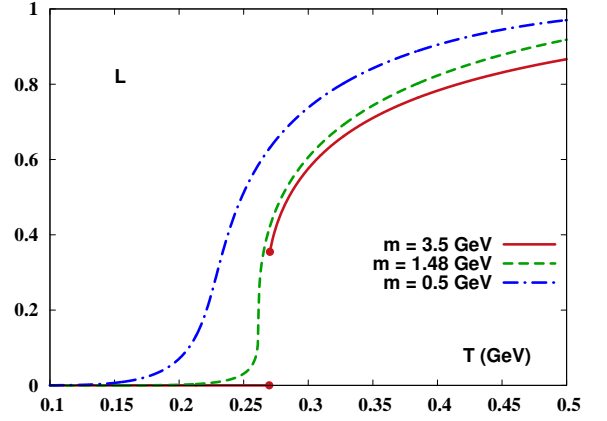}
 \caption{Temperature dependence of the thermal averaged  Polyakov loop for quark masses $m=0.5, 1.48$ and $3.5$ GeV at $N_f = 3$. }
  \label{fig:zero_mu_pl}
\end{figure}

The quark contribution to the mean field potential is obtained from (\ref{quark1}),

\begin{align}
\label{quarkf}
{U}_Q = - 2 N_f \beta^4\int \frac{d^3 {k}}{(2 \pi)^3} \, [ g^+ +  g^- ],
\end{align}

\noindent 
where the vacuum term has been dropped, since  it is a constant, independent of the Polyakov loop. 

The thermodynamic potential in (\ref{MF}) is the effective model for exploring the thermodynamics of gluons in the presence of heavy quarks.

\begin{figure*}[t]
 \includegraphics[width=3.375in]{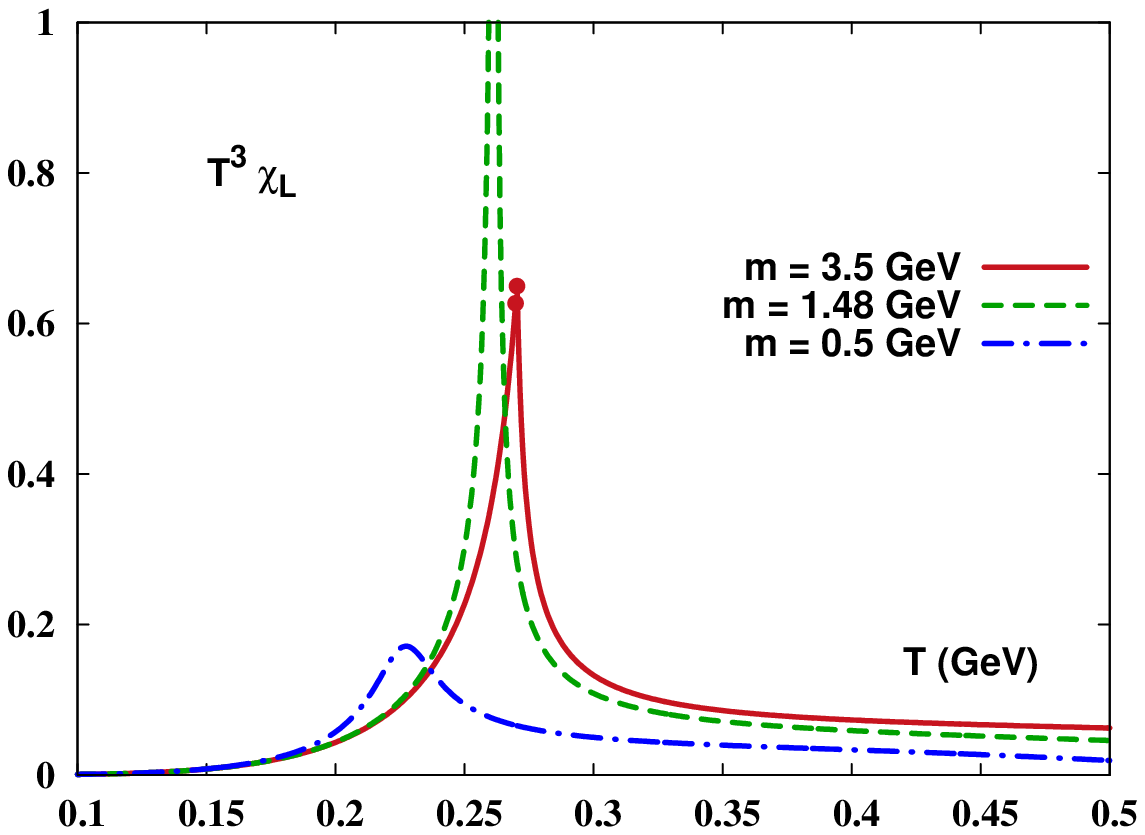}
 \includegraphics[width=3.375in]{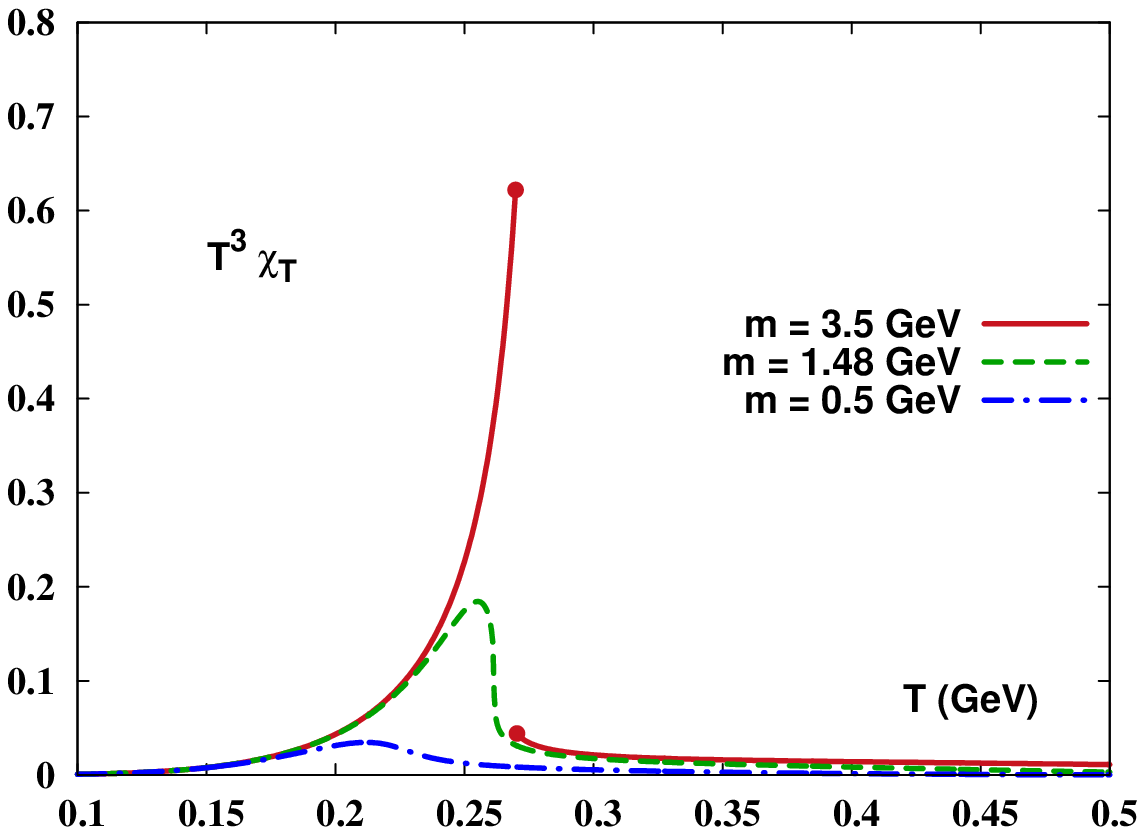}
 \caption{The temperature dependence of the longitudinal $\chi_{\rm L}$ and the transverse $\chi_{\rm T}$ Polyakov loop susceptibility for $N_f = 3$ and quark masses,  $m = 0.5, 1.48, 3.5  \, {\rm GeV}$.}
  \label{fig:zero_mu_susL}
\end{figure*}

\subsection{Heavy quarks and deconfinement }

The first-order nature of the deconfinement phase transition in the SU(3) pure gauge theory is directly related to the global Z(3) center symmetry and its spontaneous breaking. This transition is eventually washed out by the explicit symmetry breaking induced by the finite quark mass. These features are clearly exhibited in the mean field model. Indeed, in the limit of large quark masses, the leading order term in $U_Q$ in Eq.\eqref{quark1} is linear in the Polyakov loop. One finds, for $\mu=0$,

\begin{align}
        \label{h0}
        {U}_Q(L,\bar L) &\approx - h(m,N_f,T) \, L_{\rm L}, 
        \end{align}

\noindent where $L_{\rm L}$ is the thermal average of the longitudinal Polyakov loop \footnote{In the real sector of the Z(3) target space, the longitudinal  $L_{\rm L}$ and transverse $L_{\rm T}$ parts of the Polyakov loop correspond to the real and imaginary parts respectively.}, and

\begin{align}
        \label{linearb}
               h(\beta m, N_f) = \frac{6 N_f}{\pi^2T^3} \int d {k} \, {k}^2 e^{-{\beta E}},
\end{align}

\noindent is a dimensionless function of $N_f$ and $\beta m$.
 
 Equation (\ref{h0}) clearly  shows that finite mass quarks generate an external field for the Polyakov loop that breaks the Z(3) symmetry explicitly. Clearly, in the limit $\beta m \rightarrow \infty$, the strength of this field vanishes and the thermodynamics of pure gauge theory is recovered.

The strength of the symmetry breaking term increases with decreasing quark mass, eventually turning the first-order deconfinement transition into a crossover. Consequently, there is a critical value of the quark mass, $m_{\rm CEP}$, where the first-order transition ends at a second order critical endpoint \cite{GK}.

Using the linear symmetry breaking term (\ref{h0}), the location of the second order CEP is determined by \cite{PhysRevD.85.114029}

\begin{align}
        \label{phaseboundary}
        h((\beta m)_{\rm CEP},N_f) &= h_c,
\end{align}

\noindent where $h_c$ is a dimensionless constant, to be determined. The value of $h_{c}$ depends only on the Polyakov loop potential $U_{G}$. Since $h(\beta m)$ in (\ref{h0}) is a decreasing function of $\beta m$, Eqs.\eqref{linearb} and \eqref{phaseboundary} imply that the ratio $m_{\rm CEP}/T_{\rm CEP}$ increases with $N_f$. Naturally, such a pattern is also observed when the full one loop quark potential (\ref{quarkf}) is employed.

Details of the phase structure of the deconfinement transition are revealed by examining the Polyakov loop fluctuations. In the following, we study how the position of the CEP changes with the number of quark flavors at vanishing and at finite quark density.

\subsection{The deconfinement critical endpoint at $\mu=0$}

At $\mu=0$, the thermal average of the transverse Polyakov loop $L_{\rm T}$ vanishes, due to the symmetry of the partition function (\ref{main_partition}). Consequently, only  the longitudinal Polyakov loop $L_{\rm L}$ serves as an order parameter for deconfinement. On the other hand, both the longitudinal and transverse fluctuations of the Polyakov loop are non-vanishing. We introduce the longitudinal $\chi_{\rm L}$  and transverse $\chi_{\rm T}$ susceptibilities

\begin{align}
        \label{long_and_trans_1}
        \chi_{\rm L,T} = \frac{1}{2} V \, ( \langle L \bar{L} \rangle_c \pm \langle L L \rangle_c ),
\end{align}

\noindent where $\langle \dots \rangle_c$ denotes the connected part. The two terms on the r.h.s of Eq.(\ref{long_and_trans_1}) are obtained by taking the appropriate field derivatives  \cite{pmlo2, Sasaki:2006ww} of the thermodynamic potential (\ref{MF}).

The influence of heavy quarks on the deconfinement transition is clearly reflected by the properties of the Polyakov loop. In Fig.\ \ref{fig:zero_mu_pl} we show the Polyakov loop as a function of temperature for three values of the quark mass.  The expectation value of the Polyakov loop, $\langle L\rangle$ is determined by the position of the global minimum of the   potential (\ref{MF}), including the complete  one loop quark contribution (\ref{quarkf}) for three degenerate quark flavors.

For a sufficiently large quark mass, the first-order nature of the phase transition persists, while at smaller quark masses, the explicit symmetry breaking is stronger and the transition is of the crossover type. The endpoint of the line of first-order transitions defines the critical value of the quark mass, $m_{\rm CEP}$.

To identify the CEP, we consider the longitudinal fluctuations of the Polyakov loop. At the CEP,  the longitudinal  susceptibility  $\chi_{\rm L}$ diverges whereas the transverse susceptibility $\chi_{\rm T}$ remains  finite.

In Fig.\ \ref{fig:zero_mu_susL} we show the longitudinal and transverse susceptibility for three degenerate quark flavors. While both susceptibilities depend on the value of the quark mass, only the longitudinal one shows enhancement near the CEP. The transverse susceptibility decreases monotonically with decreasing quark mass. Thus, for a given $N_f$, the CEP can be located by identifying the global maximum of $\chi_{\rm L}$. Our results for the critical quark masses obtained  for different number of quark flavors at the CEP  are  as  follows,

\begin{equation}\label{mass}
        m_{\rm CEP} = 1.10, ~1.35,~ 1.48 \, {\rm GeV},~~
        {\rm for \,} N_f = 1, 2, 3.
\end{equation}


\noindent The resulting trend, with $m_{\rm CEP}$ increasing with $N_f$,  is consistent with previous findings \cite{PhysRevD.85.114029}. However, we obtain a lower value of the critical quark mass than that found in the matrix model, $m_{\rm CEP} \simeq 2.5 \, {\rm GeV}$ at $N_f=3$. As discussed in Ref. \cite{PhysRevD.85.114029}, the location of the deconfinement critical endpoint is very sensitive to the form of the Polyakov loop potential. In the present calculation,  $U_G$ reproduces the lattice data on the equation of state as well as on the susceptibilities  of the Polyakov loop. This feature is crucial for locating the CEP,  which is influenced by fluctuations of the order parameter.

\begin{figure}[t]
 \includegraphics[width=3.375in]{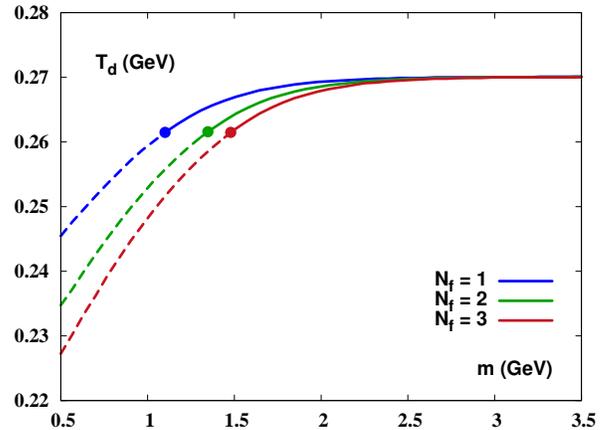}
 \caption{Quark mass dependence of the deconfinement temperature, defined by a peak of $\chi_{\rm L}$, for different $N_f$. Full lines correspond to the first-order phase transitions, while dashed lines represent the pseudo-critical temperature of the crossover transition. The points indicate the locations of the CEP.}
  \label{fig:zero_mu_Td}
\end{figure}

At a given value of the quark mass, we identify the deconfinement transition with the location of the maximum of $\chi_{\rm L}$.
In Fig.\ \ref{fig:zero_mu_Td} we show the resulting phase diagram in the $(T,m)$--plane for different $N_f$. We observe that the temperature of the CEP remains almost constant at $T_{\rm CEP} \simeq 0.261 \, {\rm GeV}$ for all $N_f$. This value is  lower than the critical temperature in the pure gauge theory by about $9 \, {\rm MeV}$. Here, the temperature obtained in the matrix model is essentially zero \cite{PhysRevD.85.114029}.

The very weak dependence of $T_{\rm CEP}$ on $N_f$, seen in Fig.\ \ref{fig:zero_mu_Td}, can be qualitatively understood. Assuming the structure of the thermodynamic potential introduced in Eq.(\ref{MF}), and keeping only the leading symmetry breaking term, $U_Q \approx -h L_{\rm L}$, the conditions for the CEP are

\begin{align}
        \label{cep_model}
        \frac{\partial {U}_G}{\partial L_{\rm L}} = h, ~~
        \frac{\partial^2 {U}_G}{\partial L_{\rm L}^2} = 0, ~~{\rm and} ~~
        \frac{\partial^3 {U}_G}{\partial L_{\rm L}^3} = 0.
\end{align}

\noindent The first condition in Eq.\eqref{cep_model} is just the gap equation, while the second and the third conditions reflect the fact  that at the CEP,  three extrema of the effective potential merge. The solution of Eq.\eqref{cep_model} fixes the critical values of the Polyakov loop, the quark mass and the temperature, $L_{\rm CEP}, m_{\rm CEP},$ and  $T_{\rm CEP}$. If ${U}_G$ is independent of the quark mass, the last two conditions in \eqref{cep_model} uniquely determine $T_{\rm CEP}$ and $L_{\rm CEP}$. Thus, in the heavy quark limit, where the leading symmetry breaking term (\ref{h0}) dominates, the critical temperature of the CEP, $T_{\rm CEP}$, 
 is independent of $N_f$.

A dependence of the effective Polyakov loop potential on the quark mass $m$ naturally appears when the complete one loop quark contribution in Eq.(\ref{quark1}) is used. The fact,  that $T_{\rm CEP}$ is almost $N_f$ independent, as shown in Fig.\ \ref{fig:zero_mu_Td}, confirms the expectation that the leading term in $U_{Q}$, shown in Eq.\eqref{h0}, is sufficient in the mass range considered.

\subsection{Phase diagram at finite chemical potential}

At finite $\mu$, the expectation values of the Polyakov loop $ L$,  and its conjugate $\bar{L}$ are both real, but in general,  different \cite{Sasaki:2006ww, PhysRevD.73.014019, PhysRevD.72.065008}. This is because at non-zero $\mu$ the effective action is complex \cite{Sasaki:2006ww, PhysRevD.72.065008}.

The expectation values of $L$ and $ \bar{L} $ are determined by solving the gap equations:

\begin{align}
        \label{finite_mu_gap}
        \frac{\partial ~\Omega}{\partial L} = 0,~~ {\rm and }~~
        \frac{\partial \Omega}{\partial \bar{L}} = 0,
\end{align}

\noindent with the thermodynamic potential $\Omega$ given in Eq.(\ref{MF}).

\begin{figure}[t]
 \includegraphics[width=3.375in]{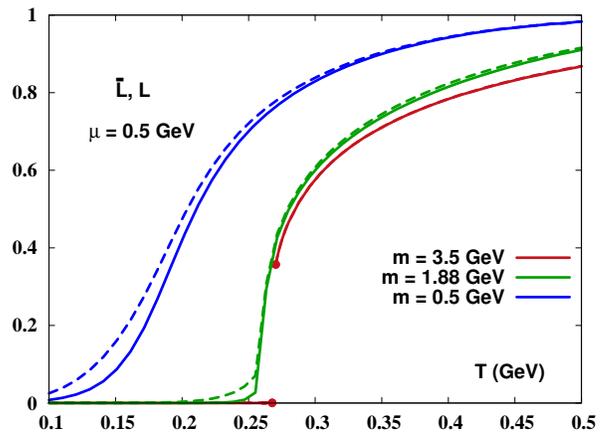}
 \caption{The expectation values of the Polyakov loop  $L$ (full lines) and its conjugate $\bar L$  (dashed lines) at $\mu=0.5 \, {\rm GeV}$, and for different values of the quark mass, $m = 3.5, 1.88, 0.5  \, {\rm GeV}$. (from right to left) }
\label{fig:finite_mu_pl}
  \end{figure}

\begin{figure*}[t]
 \includegraphics[width=3.375in]{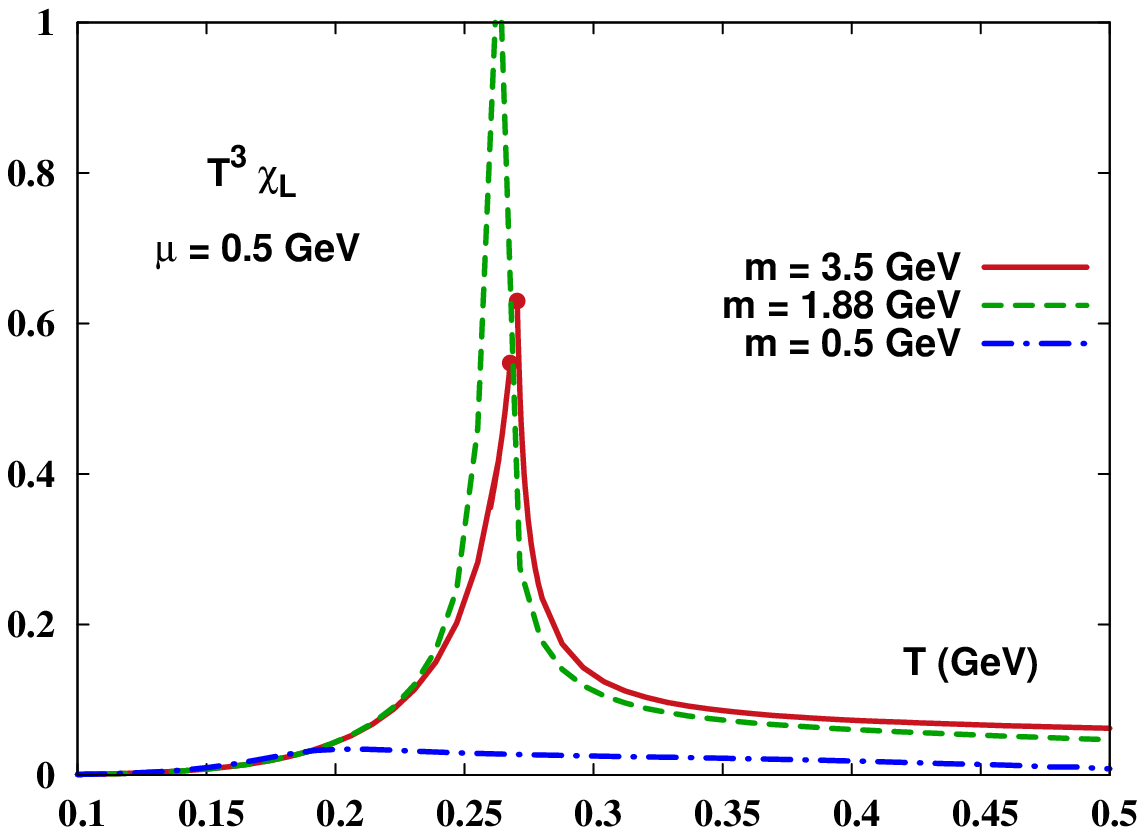}
 \includegraphics[width=3.375in]{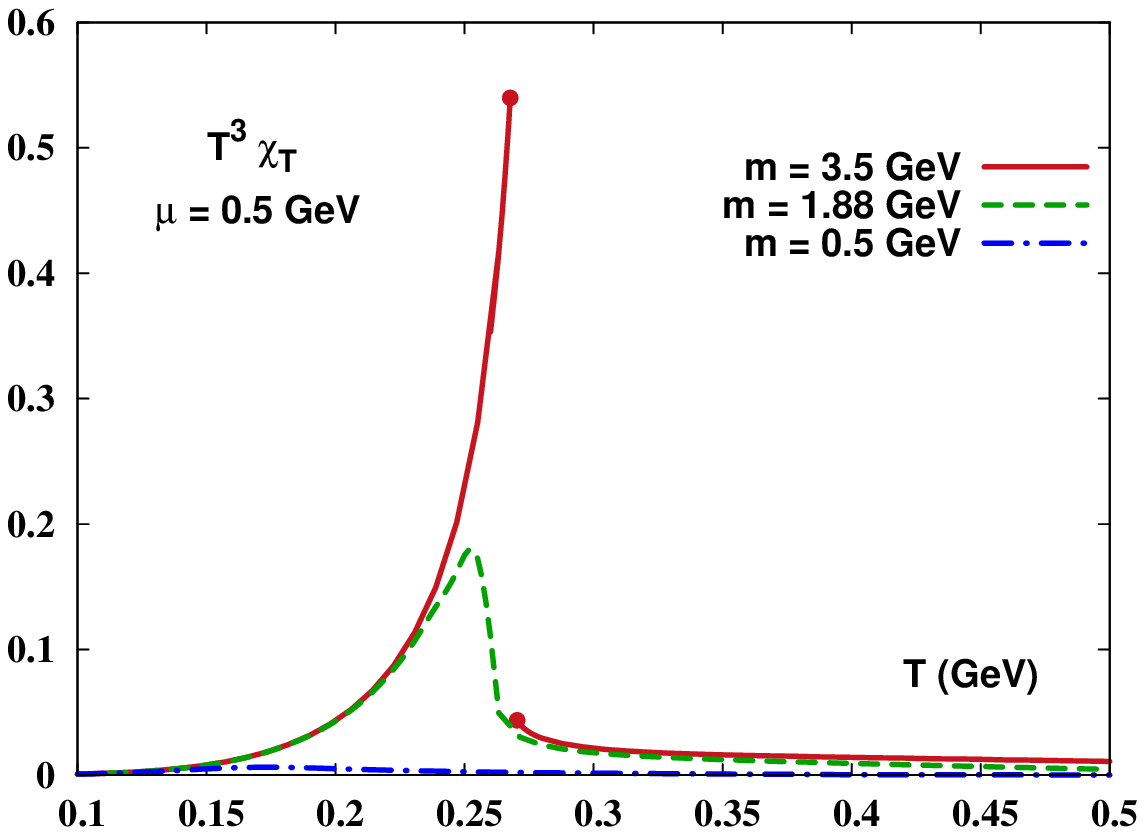}
 \caption{The longitudinal  $\chi_{\rm L}$  and transverse $\chi_{\rm T}$ susceptibility \eqref{long_and_trans_2}, versus $T$ at $\mu=0.5 \, {\rm GeV}$, and for different quark mass values.}
  \label{fig:finite_mu_susL}
\end{figure*}

In Fig.\ \ref{fig:finite_mu_pl} we show the temperature dependence of $L$ and $\bar L$ for finite  $\mu$, and for several values of the quark mass with $N_f=3$. While $ \bar{L}$ and $ L $ differ below the deconfinement transition, they merge at high temperatures. This follows from the restriction of $(L_{\rm L},L_{\rm T})$ target region,  imposed by the Haar measure. In the deconfined phase, as $L_{\rm L} \rightarrow 1$, the target region enforces $L_{\rm T} \rightarrow 0$, and thus the difference between $ \bar{L} $ and $ L $ vanishes. Such a feature is absent in the polynomial type potential, which does not comply with the constraints of the SU(3) color group structure  \cite{Sasaki:2006ww}.

The location of the CEP at finite density is again indicated by a divergence of the longitudinal susceptibility. However,   at finite $\mu$, the definitions of the longitudinal and transverse  susceptibilities are modified  \cite{Sasaki:2006ww}

\begin{align}
        \label{long_and_trans_2}
        \chi_{\rm L,T} &= \frac{1}{2} V \, [\langle {L \bar{L}} \rangle_c  \pm \frac{1}{2} \langle ({LL} + {\bar{L}\bar{L}}) \rangle_c].
\end{align}

\noindent For $\mu\to 0$,  $\langle{\bar{L}\bar{L}} \rangle \to  \langle{LL} \rangle$,  and the above definitions connect  smoothly to that, introduced in    Eq.({\ref{long_and_trans_1}) for vanishing density.

The finite density results for the longitudinal and transverse susceptibilities are shown in Fig.\ \ref{fig:finite_mu_susL}. As for $\mu = 0$, only the longitudinal susceptibility is enhanced near the CEP, whereas the transverse susceptibility is insensitive to criticality.

To construct the phase diagram, we trace the peak location of $\chi_{\rm L}$ as a function of $T$, $\mu$ and $m$. In Fig.\ \ref{fig:finite_mu_Td} we show the transition temperature as a function of $\mu$ at fixed quark mass $m = 1.6 \, {\rm GeV}$. This value is larger than
 the critical mass  found at $\mu=0$, $m_{\rm CEP}\simeq 1.48 \, {\rm GeV} $.
  Hence, in this case,  the transition is first-order at small $\mu$.

As the quark chemical potential increases, the strength of 
the explicit breaking of the Z(3) symmetry
 increases and the transition turns into a crossover. These two regimes are connected by the deconfinement CEP, located at the critical chemical potential $\mu_{\rm CEP}$. Clearly, for $m < 1.48 \, {\rm GeV}$, the system is in the crossover regime for any value of $\mu$.

The critical temperature $T_{\rm CEP}\simeq 0.261  \, {\rm GeV}$,  remains close to its $\mu = 0$ value. This follows from the assumption that $U_{G}$ is not renormalized by quark loops, which, as discussed above, is reasonably well justified in the parameter range of interest.  As $m$ increases, the CEP moves to larger $\mu$, while the critical temperature remains almost constant.

In Fig.\ \ref{fig:finite_mu_cep} we show the dependence of the critical quark mass $m_{\rm CEP}$ on the chemical potential $\mu$. The data points in the figure are extracted from the divergence of $\chi_{\rm L}$, while the line is determined using the leading term in the heavy quark limit, as discussed below. The increase of the critical quark mass with $\mu$, indicates that the first-order region shrinks with increasing density. This finding is consistent with the lattice results presented in Ref. \cite{Ejiri:2012wp, Langelage:2009jb}.

\begin{figure}[t]
 \includegraphics[width=3.375in]{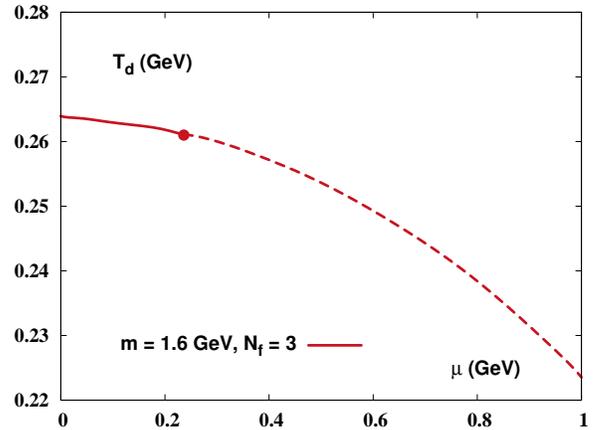}
 \caption{The $\mu$ dependence of the deconfinement temperature, defined by the peak of $\chi_{\rm L}$, for $m = 1.6 \, {\rm GeV}$ and $N_f = 3$. The full line corresponds to a first-order phase transition, while the dashed line represents the pseudo-critical temperature of the crossover transition. The point indicates the location of the CEP.}
  \label{fig:finite_mu_Td}
\end{figure}

The relation between $m_{\rm CEP}$ and $\mu_{\rm CEP}$, shown in Fig.\ \ref{fig:finite_mu_cep}, can also be studied using the leading contribution to $U_Q$. At finite $\mu$,  one finds

\begin{align}
        \label{finite_mu_h}
        {U}_Q \simeq - h(\beta\mu,\beta m,N_f) L_{\rm L},
\end{align}

\noindent where $L_{\rm L}=(L+\bar L)/2$ is the longitudinal Polyakov loop \footnote{We have neglected a subleading term  in $U_Q$, which is linear in the transverse Polyakov loop, $L_{\rm T}=(\bar L-L)/2$,  since $L_{\rm T}/L_{\rm L} \ll 1$ for $T > 0.2 \, {\rm GeV}$. (see Fig.\ \ref{fig:finite_mu_pl})}. For $m/T \gg 1$, the strength of the symmetry breaking parameter is then given by

\begin{align}\label{cosh}
 h(\beta m, \beta\mu, N_f) \simeq \frac{6 N_f}{\pi^2} (\beta {m})^2 K_2(\beta {m})\cosh(\beta {\mu}),
\end{align}

\noindent where $K_n$ is the modified Bessel function of the second kind.

As discussed above, the phase boundary is determined by 
\begin{align}\label{critmu}
        h(\beta m,\beta\mu,N_f) = h_c.
\end{align}

\noindent Since $h_c$ is a universal constant, depending only on the effective Polyakov loop potential, it can be determined at $\mu=0$. Plugging in the critical quark mass  $m_{\rm CEP} = 1.48 \, {\rm GeV}$,  the  critical temperature $T_{\rm CEP} = 0.261 \, {\rm GeV}$ for $N_f = 3$  and $\mu =0$, one finds,  $h_c \approx 0.17$.

In the parameter range considered, the critical temperature at the deconfinement CEP is both $\mu$ and $N_f$ independent, and coincides with $T_{\rm CEP}$ at $\mu=0$. Thus, Eq.(\ref{critmu}) defines the relation between the critical quark mass  and  chemical potential at the CEP

\begin{align}
        \label{finite_mu_cep_formula}
        \mu_{\rm CEP}= T_{\rm CEP} \cosh^{-1}( h_c/(\frac{6 N_f}{\pi^2} \left(((\beta {m}_{\rm CEP})^2 K_2(\beta {m}_{\rm CEP}))\right).
\end{align}

The solution of the above equation for $N_f=3$ is shown in Fig.\ \ref{fig:finite_mu_cep}. One finds a very satisfactory agreement between $m_{\rm CEP}(\mu)$ obtained from the global maximum of $\chi_{\rm L}$ and the approximate result from Eq.(\ref{finite_mu_cep_formula}).  This indicates, that to quantify the heavy-quark phase diagram at finite density, it is sufficient to retain the leading  term \eqref{finite_mu_h} in the fermion thermodynamic potential.  

\section{Model predictions and lattice results}

In constructing the effective model we assume that the parameters in the gluonic potential $U_G$ are unaffected by the presence of heavy quarks, and that the coupling of quarks to the Polyakov loop is described by the one loop potential. It is important to note, however, that polarization corrections to the transverse gluons is a potential source of $m$ dependence of the parameters of $U_G$. This is because,  in the Polyakov loop potential the transverse gluons are integrated out. If this effect can be neglected, the CEP transition temperature will remain approximately $N_f$ and $\mu$ independent. However, one expects that the dressing with fermion loops is important for light quarks. It is therefore interesting to validate these assumptions with first principle calculations on the lattice \cite{whot, PhysRevD.60.034504}.

Various lattice studies are available for the deconfinement phase transition of heavy flavors \cite{whot, PhysRevD.60.034504, Langelage:2009jb, Karsch2001579}. However, the extrapolation to the continuum has not yet been done,  and hence,  a direct comparison to  effective model is uncertain. Nevertheless, some valuable conclusions can be drawn from a comparison of the two approaches.

On the lattice, the fermionic determinant is usually expanded in terms of the lattice hopping parameter $\kappa$. The leading term in this expansion reads \cite{whot}

\begin{align}
        \ln {\det [ \hat{Q}_F (\kappa) ]} &\simeq  (2 N_f) (2 N_c) (2 \kappa)^{N_\tau} N_s ^3 \, L_{\rm L}.\label{lgt}
\end{align}

\noindent A comparison of this equation in  the mean-field approximation with the corresponding expression in the continuum model 

\begin{align}
        \ln {\det [ \hat{Q}_F ]}\simeq V T^3 h(\beta m,N_f) L_{\rm L},
\end{align}

\noindent 
using $VT^3=(N_s/N_\tau)^3$,  yields a relation between the symmetry breaking parameter $h$ in the continuum and the hopping parameter $\kappa$ on the lattice

\begin{align}\label{kappa}
        h(\beta m,N_f) \simeq (2 N_f) (2 N_c) (2 \kappa)^{N_\tau} N_\tau^3 .
\end{align}

\noindent In particular, for $\mu = 0$, one finds

\begin{align}
        \label{kappam}
 (2 \kappa)^{N_\tau} N_\tau^3=    \frac{(\beta {m})^2}{2\pi^2}  K_2(\beta {m}).
\end{align}

\noindent The right hand side of Eq.(\ref{kappam}) represents a quantity in the continuum. The continuum extrapolation of the left hand side of Eq.(\ref{kappam}) is however complicated, owing to the renormalization of the Polyakov loop and the bare quark masses represented by $\kappa$,  in Eq.(\ref{lgt}). Nevertheless, Eq.(\ref{kappam}) indicates that $(2 \kappa)^{N_\tau} N_\tau^3$  is presumably better suited for the continuum extrapolation, than the frequently used $ (2\kappa)^{N_\tau}$ combination. Indeed, the lattice calculation in Ref. \cite{PhysRevD.60.034504},  yields a very strong $N_\tau$ dependence of $(2 \kappa)^{N_\tau}$, consistent with the $N_\tau^3$ scaling implied by Eq.(\ref{kappam}). Thus, Eq.(\ref{kappam}) can be considered as an alternative prescription for relating the hopping parameter to the quark mass.

\begin{figure}[t]
 \includegraphics[width=3.375in]{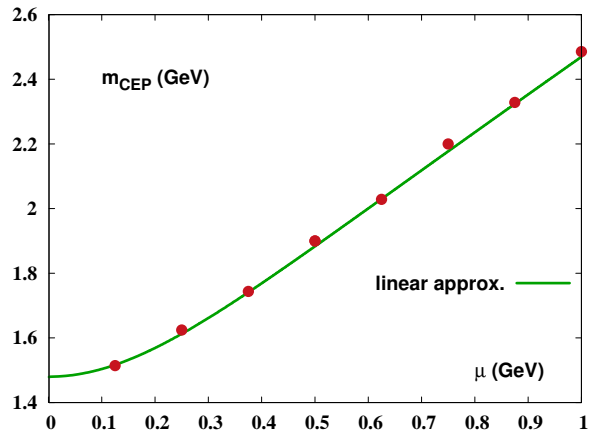}
 \caption{The critical quark mass as a function of the quark chemical potential for $N_f = 3$. Solid points represent results obtained from the global maximum of $\chi_{\rm L}$, while the line is the solution of the   Eq.\eqref{finite_mu_cep_formula}, with $h_c=0.17$ and $T_{\rm CEP}=0.261 \, {\rm GeV}$}
  \label{fig:finite_mu_cep}
\end{figure}

However, the connection of $\kappa$ to the physical quark mass remains ambiguous. Various conversion formulae have been proposed to determine the value of $m_{\rm CEP}$ \cite{DeGrand,PhysRevD.60.034504}, showing that $1<m_{\rm CEP}< 1.5 \,  {\rm GeV}$ for $N_f = 3$. The present model calculation is consistent with this mass range.

A less ambiguous extraction of the physical quark masses at the deconfinement CEP can be performed by computing the ratio of the pseudoscalar to vector meson masses, $m_{\rm PS}/m_{\rm V}$. This ratio is expected to be close to unity if the heavy quark assumption is valid. An estimate of the critical quark mass $m_{\rm CEP}$ on the lattice can then be inferred from the heavy quarkonium spectrum. This, in conjunction with the matching formula (\ref{kappam}), provides a consistency check for the continuum extraction of the critical strength of the symmetry breaking term.

\section{Conclusions}

The dependence of the SU(3) deconfinement transition in the presence of heavy dynamical quarks on the number of flavors and on the quark mass was studied within an 
effective theory. We have formulated the thermodynamic potential of interacting heavy quarks and gluons in a mean-field approach, which reproduces the equation of state and  the fluctuations of the Polyakov loop in  SU(3) pure gauge theory.}

We have explored the phase diagram at finite temperature and density for different quark masses and number of flavors. In order to examine the deconfinement critical endpoint (CEP) and its characteristics, we have studied fluctuations of the  Polyakov loop.

It was shown, that the CEP can be uniquely identified by a singularity of the longitudinal Polyakov loop susceptibility. The transverse susceptibility, on the other hand, remains finite at the CEP, and is much smaller than  at   the first-order deconfinement transition.

At vanishing chemical potential, the critical endpoint appears at the quark  mass,   $m_{\rm CEP} = 1.10, ~1.35,~ 1.48 \, {\rm GeV},$ for $N_f=1,2$ and 3 flavors, respectively.  The corresponding critical  temperature, $T_{\rm CEP}=0.261 \, {\rm GeV}$  is $9 \,  {\rm  MeV}$  below  the deconfinement temperature in the pure SU(3) gauge theory.

The critical mass, $m_{\rm CEP}$, was shown to be increasing with $\mu$,  whereas  the critical temperature at the CEP is almost constant. Consequently, at finite density,  the region of the first-order deconfinement phase transition shrinks with increasing $\mu$.

We have discussed the relation between the model and lattice results. In particular, we have argued, that in order to obtain the continuum limit for the strength of the $Z(3)$ symmetry breaking term in the hopping parameter ($\kappa$) expansion, one should consider the product, $(2 \kappa)^{N_\tau} N_\tau^3$,  where $N_\tau$ is the lattice size in the temporal direction.

Finally, based on the matching criteria of the model and the lattice fermionic determinant, a formula connecting the hopping parameter and the quark mass   was  proposed.

\

\section{Acknowledgments}

We acknowledge the stimulating discussions with Chihiro Sasaki, the BNL-Bielefeld lattice group, and  the WHOT-QCD lattice Collaboration. P.M.L is grateful for the helpful comments from Constantia Alexandrou. K. R.  acknowledges fruitful discussions with Frithjof  Karsch and Rob  Pisarski. P.M.L. is supported by the Frankfurt Institute for Advanced Studies (FIAS). B. F. is supported in part by the Extreme Matter Institute EMMI. K. R. acknowledges partial support of the Polish National Science Center (NCN), under Maestro grant  2013/10/A/ST2/00106. The numerical calculations have been performed on the GridEngine Cluster at GSI.


\begin{thebibliography}{10}

\bibitem{Greensite}
  J.~Greensite,
  Prog.\ Part.\ Nucl.\ Phys.\  {\bf 51}, 1 (2003).

\bibitem{tHooft}
  G.~'t Hooft,
  Nucl.\ Phys.\ B {\bf 138}, 1 (1978).

\bibitem{Yaffe}
  L.~G.~Yaffe and B.~Svetitsky,
  Phys.\ Rev.\ D {\bf 26}, 963 (1982).


\bibitem{McLerran}
  L.~D.~McLerran and B.~Svetitsky,
  Phys.\ Lett.\ B {\bf 98}, 195 (1981);
  Phys.\ Rev.\ D {\bf 24}, 450 (1981).

\bibitem{Polyakov}
  A.~M.~Polyakov,
  Phys.\ Lett.\ B {\bf 72}, 477 (1978).

\bibitem{Binder}
  K.~Binder,
  Rep.\ Prog.\ Phys.\ {\bf 50}, 783 (1987).

\bibitem{GK}
F.~Green and F.~Karsch,
Nucl. \ Phys. \ B {\bf 238}, 297 (1984).

\bibitem{pmlo1}
P.~M.~Lo, B.~Friman, O.~Kaczmarek, K.~Redlich, and C.~Sasaki,
Phys.\ Rev.\ D {\bf 88}, 014506 (2013).

\bibitem{2+1QCD}
  C.~McNeile, A.~Bazavov, C.~T.~H.~Davies, R.~J.~Dowdall, K.~Hornbostel, G.~P.~Lepage and H.~D.~Trottier,
  Phys.\ Rev.\ D {\bf 87}, 034503 (2013).

\bibitem{pmlo2}
P.~M.~Lo, B.~Friman, O.~Kaczmarek, K.~Redlich, and C.~Sasaki,
Phys. \ Rev. \ D {\bf 88}, 074502 (2013).



 \bibitem{Karsch2001579}
F.~Karsch, E.~Laermann, and A.~Peikert,
Nucl. \ Phys. \ B {\bf 605},  579 (2001).


\bibitem{DeGrand}
  T.~A.~DeGrand and C.~E.~DeTar,
  Nucl.\ Phys.\ B {\bf 225}, 590 (1983).


\bibitem{attig}
N. Attig, B. Petersson, M. Wolff, R.V. Gavai
Z. Phys. C - Particles and Fields  {\bf 40},  471 (1988).

\bibitem{Langelage:2009jb}
  J.~Langelage and O.~Philipsen,
  JHEP {\bf 1001}, 089 (2010).

\bibitem{Ejiri:2012wp}
S.~Ejiri {\it et al.}  [WHOT-QCD Collaboration],
PoS LATTICE {\bf 2012}, 089 (2012).


  \bibitem{PhysRevD.60.034504}
  C.~Alexandrou, A.~Borici, A.~Feo, P.~de Forcrand, A.~Galli, F.~Jergerlehner and T.~Takaishi,
  Phys.\ Rev.\ D {\bf 60}, 034504 (1999).

  \bibitem{Aarts:2001dz}
  G.~Aarts, O.~Kaczmarek, F.~Karsch and I.~-O.~Stamatescu,
  Nucl.\ Phys.\ Proc.\ Suppl.\  {\bf 106}, 456 (2002).

 \bibitem{fomm}
  M. Fromm, J. Langelage, S. Lottini, O. Philipsen
  JHEP {\bf 01},  042 (2012).



\bibitem{whot}
H.~Saito, S.~Ejiri, S.~Aoki, T.~Hatsuda, K.~Kanaya, Y.~Maezawa, H.~Ohno, and
  T.~Umeda,
Phys. \ Rev. \ D {\bf 84}, 054502 (2011) .



\bibitem{PhysRevD.85.114029}
  K.~Kashiwa, R.~D.~Pisarski and V.~V.~Skokov,
  Phys.\ Rev.\ D {\bf 85}, 114029 (2012).


\bibitem{meisinger}
  P.~N.~Meisinger, T.~R.~Miller, and M.~C.~Ogilvie,
  Phys.\ Rev.\ D {\bf 65}, 034009 (2002) .
  P. N. Meisinger and M. C. Ogilvie, Phys. Rev. D {\bf 65} , 056013 (2002) .  P. N. Meisinger, M. C. Ogilvie and T. R. Miller, Phys. Lett. {\bf B} 585 (2004) 149.


\bibitem{gross}
D. J. Gross, R. D. Pisarski and L. G. Yaffe, Rev. Mod. Phys. {\bf 53}  (1981) 43.


\bibitem{kr}
  H.~T.~Elze, D.~E.~Miller and K.~Redlich,
  Phys.\ Rev.\ D {\bf 35}, 748 (1987).

\bibitem{PhysRevD.69.097502}
  Y.~Hatta and K.~Fukushima,
  Phys. \ Rev. \ D {\bf 69}, 097502 (2004).

\bibitem{Sasaki:2006ww}
C.~Sasaki, B.~Friman, and K.~Redlich,
Phys.\ Rev.\ D {\bf 75}, 074013 (2007) .


\bibitem{PhysRevD.73.014019}
C.~Ratti, M.~A.~Thaler, and W.~Weise,
  Phys.\ Rev.\ D {\bf 73},  014019 (2006).

\bibitem{PhysRevD.72.065008}
  A.~Dumitru, R.~D.~Pisarski and D.~Zschiesche,
  Phys.\ Rev.\ D {\bf 72}, 065008 (2005) .




\end{thebibliography}
\end{document}